\documentclass{aa}

\usepackage{amsmath,amssymb}
\usepackage{graphicx}
\usepackage{txfonts}

\begin{document}

\title{Coronal rain in magnetic bipolar weak fields}

\author{C. Xia, R. Keppens, X. Fang}

\institute{Centre for mathematical Plasma Astrophysics, Department of
Mathematics, KU Leuven, Celestijnenlaan 200B, 3001 Leuven, Belgium\\
email:~{chun.xia@kuleuven.be}}

\abstract
{}
{We intend to investigate the underlying physics for the coronal rain phenomenon in a 
representative bipolar magnetic field, including the formation and the dynamics of coronal
rain blobs. }
{With the MPI-AMRVAC code, we performed three dimensional radiative magnetohydrodynamic (MHD)
simulation with strong heating localized on footpoints of magnetic loops after a relaxation 
to quiet solar atmosphere.}
{Progressive cooling and in-situ condensation starts at the loop top
due to radiative thermal instability. The first large-scale condensation on the loop top 
suffers Rayleigh-Taylor instability and becomes fragmented into smaller blobs. The blobs fall vertically 
dragging magnetic loops until they reach low-$\beta$ regions and start to fall along 
the loops from loop top to loop footpoints. A statistic study of the coronal rain blobs finds 
that small blobs with masses of less than $10^{10}$ g dominate the population. When blobs fall
to lower regions along the magnetic loops, they are stretched and develop a non-uniform 
velocity pattern with an anti-parallel shearing pattern seen to develop along the central axis of the blobs. Synthetic
images of simulated coronal rain with Solar Dynamics Observatory Atmospheric Imaging Assembly 
well resemble real observations presenting dark falling clumps in hot channels and bright rain 
blobs in a cool channel. We also find density inhomogeneities during a coronal rain ``shower'',
which reflects the observed multi-stranded nature of coronal rain.}
{}


\maketitle

\section{INTRODUCTION}\label{intro}

Coronal rain refers to cool and dense elongated plasma blobs or thread segments, which suddenly 
appear in the low corona, falling along coronal loops all the way down to the solar surface.
This phenomenon was first recorded and classified as coronal sunspot prominences, commonly originating 
in coronal space and pouring down to spot regions \citep{Pettit43}, and was later observed 
and clarified as ``coronal rain'' by its characteristic feature of rapid brightening when 
approaching the chromosphere in the early 1970s \citep{Kawaguchi70}. It has been well observed 
in spectral lines sampling temperature ranges from transition region (TR) to chromosphere 
\citep{deGroof05,Kamio11,Antolin10,Antolin11}. High-resolution spectroscopic observations 
\citep{Antolin12} found that coronal rain is composed of elongated blobs with average widths 
of 310 km, lengths of 710 km, average temperatures below 7000 K, average falling speed of 70 
km s$^{-1}$, and accelerations much smaller than the gravitational acceleration projected
along loops. In non-flaring coronal loops, coronal rain is observed to appear during 
progressive cooling with high-speed downflow and dramatic evacuation in the loops 
\citep{Schrijver01, deGroof05}. \citet{OShea07} found spectroscopic evidence of plasma 
condensations taking place in coronal loops and forming coronal rain. \citet{Antolin12} found 
that coronal rain often occurs simultaneously in neighboring magnetic strands, forming groups 
of condensations which are seen as large clumps if they are close enough in proximity. These are called 
``showers'' and can have widths up to a few Mm. \citet{Antolin15} found that coronal rain is 
a highly multi-thermal phenomenon with the multi-wavelength emission located very closely.
The fragmentary and clumpy appearance of coronal rain blobs at coronal heights becomes more 
continuous and persistent at chromospheric heights immediately before impact. Rain clumps appear 
organized in parallel strands with density inhomogeneities in both chromospheric and TR 
temperatures. The electron density of rain clumps was found to be about 1.8-7.1$\times10^{10}$ 
cm$^{-3}$, through estimation based on absorption in multiple extreme ultraviolet (EUV) channels. Besides this 
quiescent coronal rain, which occurs in nonflaring coronal loops with relatively weak
variation of energy and mass, flare-driven coronal rain, which appears in postflare loops as 
a result of catastrophic cooling, often emerges as a bunch of parallel strands extending
 from loop top to footpoint \citep{Scullion14,Scullion16}.

With the phenomena of progressive cooling and in situ condensation, coronal rain is believed 
to be a representative of the general phenomenon of radiative thermal instability 
\citep{Parker53,Field65} in an astrophysical plasma, that occurs whenever energy losses due 
to radiation overcome the heating input and the resulting cooling further amplifies energy losses.
Numerical simulations firstly performed in one-dimensional setups \citep{Anti99,Karpen01,
Muller03,Muller04,Muller05,Xia11} have demonstrated that a heating input of a coronal 
loop concentrated at both footpoints, in chromosphere and near TR, can cause the loop to experience thermal non-equilibrium and thermal instability. This 
leads to coronal rain formation or prominence formation depending on the magnetic 
configuration of the loop. Footpoint heating of coronal loops is supported by 
observational evidence \citep{Aschwanden01b} and by hydrostatic coronal loop models 
\citep{Aschwanden01a}. Three-dimensional (3D) magnetohydrodynamics (MHD) numerical 
simulations \citep{Hansteen10} found a concentration of Joule heating toward the upper 
chromosphere, TR, and lower corona as a result of the braiding of the magnetic field 
decreasing exponentially with height. Observations 
\citep{DePontieu11} have shown a ubiquitous coronal mass supply where chromospheric plasma 
is accelerated and heated in type II spicules upward into the corona, and then leads to
fading of type II spicules in chromospheric spectral line images 
\citep{DePontieu07,Rouppe09}. 

Inspired by the first two-dimensional (2D) evaporation-condensation model 
demonstrating prominence formation in a magnetic arcade \citep{Xia12}, 
\citet{Fang13} presented the first 2D coronal rain simulation, in which a large 
zigzag shape condensation forms across the top regions of a magnetic arcade 
covering a range of several Mm, then splits into many small blobs with elongated
side wings and V-shaped features, descending along both sides of the arcade field 
lines. They performed statistic analyses to quantify blob widths and lengths, which average 
400 km and 800 km, respectively, and velocities up to 65 km s$^{-1}$, which is smaller than the observed 
average speed. Later, \citet{Fang15} extended their model with an increased resolution
of 20 km per cell and much longer time coverage, and studied the blob condensation in
detail and found recurrent coronal rain showers to occur in limit cycles. They quantified
the thermal structure of blob-corona TR and the variations of 
density, kinetic energy, and temperature during the impact of rain blobs on the 
chromosphere. They also showed how high-speed anti-parallel shear flows at two sides 
of condensations are induced that further facilitate fragmentation of the condensations.
\citet{Moschou15} performed 3D simulations on coronal 
condensation in a magnetic configuration of a quadrupolar arcade system. In their
models, coronal condensations occur in nearly horizontal coronal loop parts, where dense blobs first descend through (weak) local magnetic field undergoing Rayleigh--Taylor instability (RTI). Later on, small blobs start to follow more closely the magnetic field lines in the lower regions near the footpoints and slide down to the chromosphere, behaving
like coronal rain. Their magnetic configuration did not represent a typical solar active 
region where coronal rain preferentially occurs. In a magnetic structure built 
from an observed magnetogram, \citet{Mok16} simulated an active region validated with 
simulated EUV emissions. They found thermal non-equilibrium in a coronal loop
leading to cooling and coronal condensation, although their spatial resolution was
not enough to obtain coronal rain blobs. 

In order to better understand coronal rain in realistic coronal loops, we perform a 3D MHD 
simulation on quiescent coronal rain in a bipolar magnetic field, which is an idealized
magnetic structure of a two-sunspot solar active region. The numerical methods and the
 simulation strategy are explained in Section~\ref{sec:method}. We present 
the results in Section~\ref{sec:results} and end with a conclusion and discussion in 
Section~\ref{sec:conc}.

\section{Numerical method} \label{sec:method}
We setup a 3D Cartesian simulation box with extensions in $x$ from -20 Mm to 20 Mm, 
in $y$ from -30 Mm to 30 Mm, and in $z$ from 0 Mm to 60 Mm. The solar gravity points
along the negative $z$ direction and the $z=0$ bottom surface represents the solar surface. With 
five levels of adaptive mesh refinement, the mesh has an effective resolution of 
$384\times576\times768$, with smallest grid cell size of 78 km. To mimic a magnetic
configuration of a simple two-sunspot active region, we use a potential magnetic 
field composed of two dipoles \citep{Torok03} with vertical and oppositely directed 
moments located below the solar surface at (0, $y_c$, $-z_c$) and (0, $-y_c$, $-z_c$), 
respectively:
\begin{eqnarray*}
 \begin{aligned}
  & B_x=B_{x+}+B_{x-},~~B_y=B_{y+}+B_{y-},~~B_z=B_{z+}+B_{z-},\\
  & B_{x\pm}=3x(x+z_c)f_{\pm},\\
  & B_{y\pm}=3(y \mp y_c)(z+z_c)f_{\pm},\\
  & B_{z\pm}=[2(z+z_c)^2-x^2-(y\mp y_c)^2]f_{\pm},\\
  & f_{\pm}=\frac{\pm B_c}{[x^2+(y\mp y_c)^2+(z+z_c)^2]^{2.5}}, 
 \end{aligned}
\end{eqnarray*}
where $y_c=10$ Mm, $z_c=15$ Mm, and $B_c=200$ G. This magnetic setup gives maximal 120 G
on the bottom surface ($z=0$) and about 3 G at a height of 30 Mm. Although the magnetic
field strength is much weaker than a typical solar active region, these weak magnetic 
loops may be found in a decayed active region. 
To construct a solar atmosphere,
we let the model obtain a realistic thermal and gravitational stratification by 
relaxing an initial state, which has a vertical distribution of temperature of about 
8000 K below the TR at a height of 2 Mm with smooth connection (TR at 160,000 K 
using hyperbolic tangent function) to the upper region where the temperature 
increases with height in such a way 
that the vertical thermal conduction flux is always $2\times10^5$ erg cm$^{-2}$ s$^{-1}$. 
The density is then calculated solving a hydrostatic equation with the number 
density at the bottom being $7.3\times10^{12}$ cm$^{-3}$.
Our simulations are performed by solving the thermodynamic  
MHD equations given by
\begin{align}
 \frac{\partial \rho}{\partial t}+\nabla\cdot\left(\rho\mathbf{v}\right)&=0,\\
 \frac{\partial \left(\rho\mathbf{v}\right)}{\partial t}+\nabla\cdot\left(
 \rho\mathbf{vv}+p_{\rm tot}\mathbf{I}-\frac{\mathbf{BB}}{\mu_0}\right)&=\rho\mathbf{g},\\
 \frac{\partial E}{\partial t}+\nabla\cdot\left(E\mathbf{v}+p_{\rm tot}\mathbf{v}-
 \frac{\mathbf{BB}}{\mu_0}\cdot\mathbf{v}\right)&=\rho\mathbf{g}\cdot
     \mathbf{v}+H-R+\nabla\cdot\left(\boldsymbol{\kappa}\cdot\nabla T\right), \\
 \frac{\partial \mathbf{B}}{\partial t}+\nabla\cdot\left(\mathbf{vB}-\mathbf{Bv}\right)&=0,
\end{align}
where $\rho$, $\mathbf{v}$, $\mathbf{B}$, and $\mathbf{I}$ are the plasma
density, velocity, magnetic field, and unit tensor, respectively. Moreover,  
$p_{\rm tot}\equiv p+B^2/2\mu_0$ is total pressure with gas pressure 
$p=2.3 n_{\rm  H} k_{\rm B} T$ assuming full ionization and an approximate 
helium abundance ($n_{\rm He}/n_{\rm H}=0.1$),  
$E=p/(\gamma-1)+\rho v^2/2+B^2/2\mu_0$ is total energy, and $\mathbf{g}=-
g_\odot r_\odot^2/(r_\odot+z)^2\mathbf{\hat{z}}$ is the gravitational
acceleration with solar radius $r_\odot$ and the solar surface
gravitational acceleration $g_\odot$. We use normalization units of length $L_0=10$ Mm,
time $t_0=1.43$ min, temperature $T_0=10^6$ K, number density $n_0=10^9$ cm$^{-3}$, velocity $v_0=116.45$ km 
s$^{-1}$, and magnetic field $B_0=2$ G to normalize the equations. Any number without
a specified unit has the according normalization unit thoughout the paper. We use the 
Adaptive Mesh Refinement Versatile Advection Code (MPI-AMRVAC) 
\citep{Keppens12,Porth14} to numerically solve these equations with a scheme 
setup combining the Harten--Lax--van Leer Riemann solver \citep{Harten83} with a 
third-order slope limited reconstruction \citep{Cada09} and a three-step Runge--Kutta 
time integration. We add a diffusive term into the induction equation to keep the 
divergence of magnetic field under control \citep{Keppens03,vanderH07,Xia14}.
Thermal source terms in the energy equation, such as field-aligned 
($\boldsymbol{\kappa}=\kappa_\parallel \mathbf{bb}$) thermal conduction and
optically thin radiative cooling $R=1.2 n_{\rm H}^2\Lambda(T)$, are treated as
in our previous studies \citep{Xia14,Xia16} using a Super Time Stepping scheme 
\citep{Meyer12} and an exact integration scheme \citep{Townsend09}, respectively. 
To establish a hot corona, the global
coronal heating is simulated by adding a parametrized heating term 
$H_1=c_1 B^{1.75} n_{\rm e}^{0.125}/r^{0.75}$ erg cm$^{-3}$ s$^{-1}$
\citep{Lionello13,Mok16}, where $c_1=4.54$ is a constant, $B$ is magnetic field 
strength, $n_{\rm e}=1.2 n_{\rm H}$ is number density of electrons, and 
$r=1/|(\nabla \mathbf{b}) \cdot \mathbf{b}|$ is the local radius of curvature of 
magnetic field lines, where $\mathbf{b}=\mathbf{B}/B$ is a unit vector along the magnetic field.
We split off the potential magnetic field as a time-invariant component, and numerically solve for the deviation from it \citep{Tanaka94,Porth14} to avoid negative
pressure problems caused by errors in magnetic energy when magnetic energy dominates
the internal energy in extremely low-$\beta$ plasma. On the four side boundaries, we use
a symmetry boundary condition for density and gas pressure, and extrapolate magnetic
field according to a zero-gradient assumption. On the bottom boundary, we fix the 
density, pressure, and magnetic field the same as in the initial condition. On the top boundary, we solve the hydrostatic equation to obtain density and pressure values assuming 
a continuous temperature distribution, and adopt the same treatment for magnetic field as 
on side boundaries.  We use anti-parallel symmetry conditions for velocity to ensure zero
velocity on the all boundary surfaces. The normal component of extrapolated magnetic 
field is modified to numerically fulfill a centered difference zero-divergence constraint.

The initial setup is not in an equilibrium, so we first run the simulation with only global 
heating in a relaxation period of 114 min to reach a quasi-equilibrium. Starting from that state, we 
reset the time to zero and add a localized heating to evaporate chromospheric plasma into the
coronal loop to cool down to coronal rain. The localized heating $H_2$ is concentrated at both
foot point chromospheric regions of the arcade with strong vertical magnetic field and 
decays with height above $z=4$ Mm in a Gaussian profile:
\begin{equation*}
H_2=\begin{cases}f(t) c_2 (B_z/B_h)^2 e^{-((z-z_h)/H_m)^2}  & \text{if $z > z_h$} \\
      f(t) c_2 (B_z/B_h)^2  & \text{if $z \le z_h$},
    \end{cases}
\end{equation*}
where $c_2=10^{-2}$ erg cm$^{-3}$ s$^{-1}$, $B_h=68$ G, $z_h=4$ Mm, $H_m=3.16$ Mm, and $f(t)$
is a linear ramp function of time to switch on the heating smoothly and keep it fixed 
afterwards. This heating mimics long-term mild heating in nonflaring coronal loops, and
can be modified to represent impulsive violent heating of flares to stimulate postflare loops 
and flare-driven coronal rain in the future. This final stage is simulated for 114 min, and 
we present the results in the following sections.

\section{RESULTS}\label{sec:results}

\begin{figure*}
\centering
\includegraphics[width=0.9\textwidth]{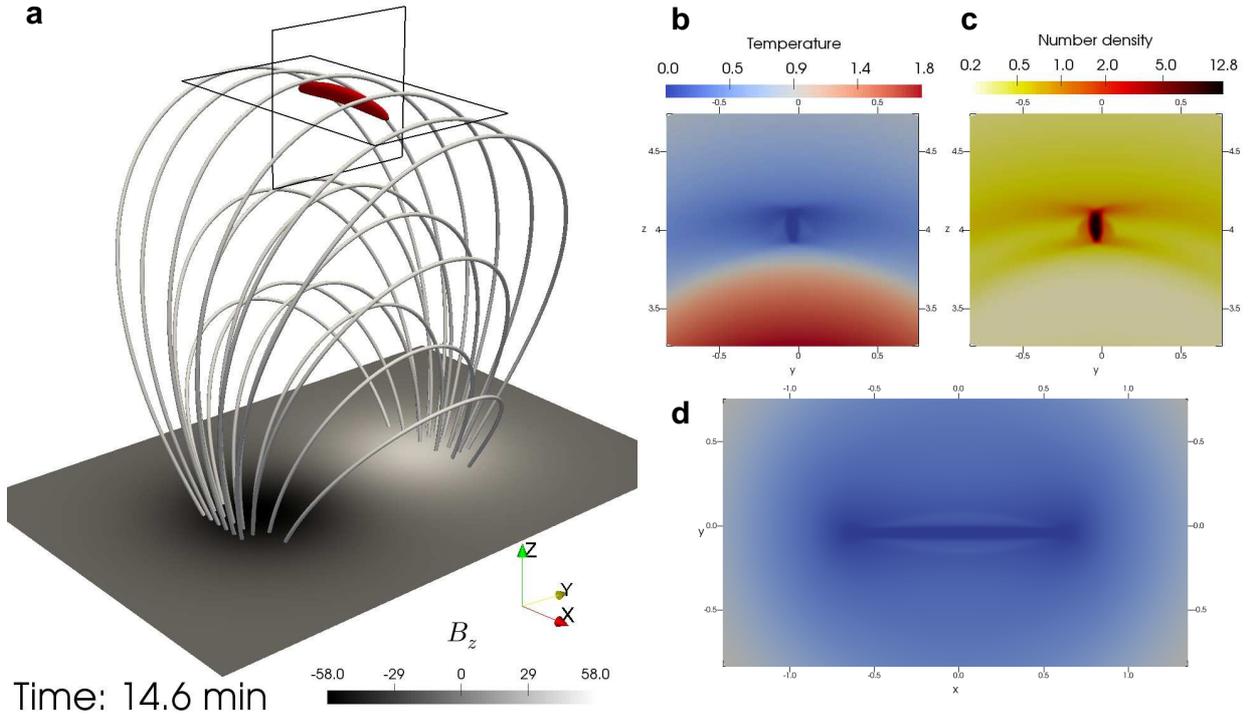}
\caption{(a) A 3D view of the first condensed blob at 14.6 min shown by a red density 
isosurface at $7\times10^{9}$ cm$^{-3}$
showing the condensed blob, magnetic field lines showing bipolar coronal loops, and bottom
plane colored by vertical magnetic field. The field lines are integrated from fixed points at
the bottom plane; A vertical slice in the $y$-$z$ plane ($x=0$, 
bounded by the vertical black frame in (a)) through the condensation showing number 
density in panel (b) and temperature in panel (c); (d) A horizontal slice in the $x$-$y$ plane ($z=40$ 
Mm, bounded by the horizontal black frame in (a)) showing temperature structures with the same 
color-scale as in panel (b). (Numbers are dimensionless with according units of time 
$t_0$, magnetic field $B_0$, temperature $T_0$, and number density $n_0$. We note that all red 
isosurfaces and bottom magnetograms in the following figures have the same values as defined 
here.}
\label{fig:first}
\end{figure*}

The localized heating evaporates chromospheric plasma into hot coronal plasma and causes 
upflows from footpoint regions with speeds peaking at about 50 km s$^{-1}$ in the lower 
corona. A loop top region centered at (0,0,40) Mm takes the lead in cooling because the
 radiative cooling there dominates over heating and thermal conduction as a result of density 
enhancement. This enhancement is caused by the evaporated flows along the heating loops and by 
squeezing due to the expansion of underlying heated loops. At 13.5 min, the central
region cools down to 0.02 MK, below which temperature our radiative cooling vanishes. Then,
induced siphon flows, with speeds up to 24 km s$^{-1}$ coming from two sides, merge there and
 create a high-density low-temperature plasma blob. This blob grows in three dimensions 
and the most rapid extension happens along the direction parallel to the polarity inversion 
line ($x$-axis) as shown in Figure~\ref{fig:first}. This is caused by the similarity and 
concurrency of the condensation process across the loops with a similar shape and height 
distributed along the $x$-direction. Observations also found that condensations in neighboring 
magnetic field strands occur simultaneously, which was thought to be due to a common footpoint 
heating process \citep{Antolin12}. The head-on impact of siphon flows not only compresses the
central cool plasma to form a dense plasma blob, but also at the same time launches two 
rebounding slow shocks propagating against the continuous siphon flows along the magnetic field. 
The shock fronts are clearly seen in the temperature and density maps on the slices through 
the condensation as shown in Figure~\ref{fig:first} (b)--(d). The shock-front surfaces
are curved, slow shock fronts and show up on a bundle of similar magnetic flux tubes where 
condensations occur successively. The processes of condensation and shock formation on these flux 
tubes are essentially the same, although they start at different times. Therefore, the 
whole history of the condensation process in a single flux tube is represented by snapshots like
Figure~\ref{fig:first} (b)--(d). Similar rebound shocks were also found in our previous
2.5D simulations \citep{Xia12,Fang15}. At this moment, the shocks decrease the inflow speed 
from about 50 km s$^{-1}$ to about 17 km s$^{-1}$, compress the plasma from $5\times10^{8}$ 
to $9\times10^{8}$ cm$^{-3}$, and heat the plasma from 0.18 MK to 0.21 MK.

\begin{figure*}
\centering
\includegraphics[width=0.9\textwidth]{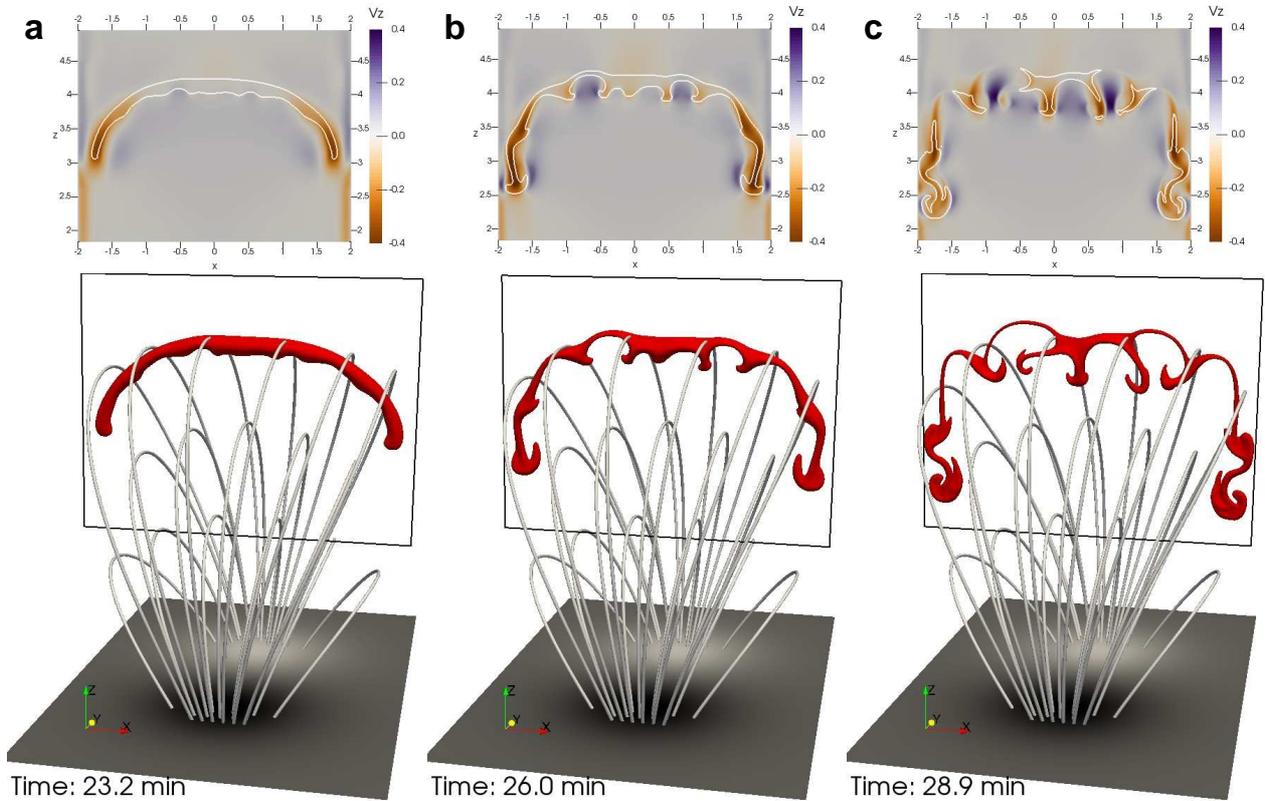}
\caption{Time series of snapshots ((a) at 23.2 min, (b) at 26.0 min, (c) at 28.9 min)
presenting the fragmentation of the large condensation due to RTI.
In the top row, vertical slices ($x$-$z$ planes), depicted by the black frames in the lower 
3D views, are colored by the vertical component of velocity with white density contours at
number density of $7\times10^{10}$ cm$^{-3}$ .}
\label{fig:RTI}
\end{figure*}

As the condensation extends with a span of 37 Mm across loop tops, it starts to distort and
becomes fragmented in about 6 min due to RTI, which is shown in 
Figure~\ref{fig:RTI} where the vertical velocity pattern alternates upflows with downflows
in a direction perpendicular to the magnetic field. The dense coronal rain plasma is gathered 
in falling spikes and upflows correspond to hot coronal plasma. The plasma $\beta$ of the
condensation is about 0.5 and the magnetic field strength is about 2 G. The Atwood number 
$\mathbf{A}$, which is $(\rho_{high}-\rho_{low})/(\rho_{high}+\rho_{low}),$ indicating the 
density disparity between two layers of fluid under acceleration in classical RTI
theory \citep{Chandrasekhar61}, for the coronal rain plasma and underlying coronal plasma is
about 0.93. The coronal rain plasma is about 25 times denser than the underlying coronal plasma.
Similar results were found in our previous studies \citep{Moschou15}. As shown in 
Figure~\ref{fig:RTI} (c), the falling spikes of coronal rain plasma have larger growth rates and 
penetrate deeper into the opposite region than rising bubbles of coronal plasma as expected for 
Atwood numbers close to 1. We note that the RTI modes in this study are almost perpendicular to local 
magnetic field, which should locally behave as in 
hydrodynamics where shortest wavelengths grow the fastest. However, since the magnetic field lines are line-tied at the bottom, the interchange deformation
of magnetic field lines and the RTI mode growth is restrained. 

\begin{figure*}
\centering
\includegraphics[width=0.9\textwidth]{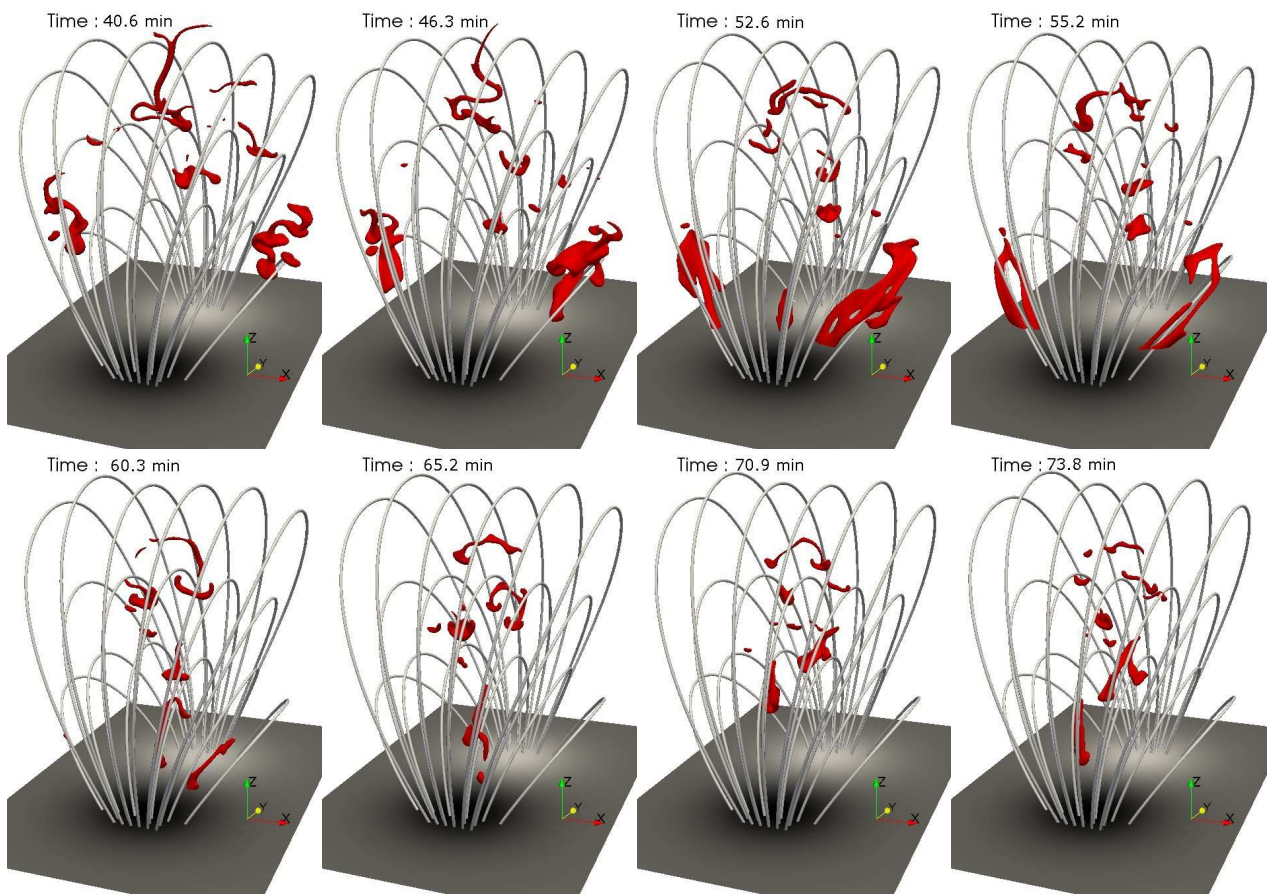}
\caption{Time series of snapshots on the evolution of coronal rain. Red density isosurfaces above a 
height of 4 Mm are plotted to show coronal rain blobs above the TR. Supplementary Movie 1 shows the 
temporal evolution of these snapshots in a different viewing angle.}
\label{fig:evolve}
\end{figure*}

The fragmented blobs move downward in three groups, namely, two large clumps at the two flanks where 
coronal loops are inclined, and one near the central vertical loops, as shown in the first row of 
Figure~\ref{fig:evolve}. These blobs initially fall roughly in the vertical plane ($y=0$) 
 along the PIL. Because the evolution of our bipolar system is roughly symmetric and magnetic 
field is weak enough at those loop top regions, the blobs, sitting at loop tops with nearly zero projected 
gravity along the loops, can displace magnetic field transversally dragging down their hosting magnetic 
loops. Blobs of the central group fall slower than those at 
flanks due to the stronger resistant magnetic-tension force they encounter. They fragment further into 
smaller ones. Many tiny blobs gradually evaporate to hot plasma and disappear in the 
density isosurface view. The flank clumps fall almost vertically with speeds about 16 km 
s$^{-1}$ until they reach a lower height of about 13 Mm. Then, they start to be more
constrained in their motions by the local magnetic field with plasma $\beta$ about 0.1 and to be forced to
stream down along the magnetic loops toward the foot points in the negative polarity region. After the flank blobs
drain down completely, coronal rain blobs mainly form at the top region of the central loops and fall
along the loops intermittently (see the second row of Figure~\ref{fig:evolve}). During their fall, the 
condensed blobs keep changing shape, elongating, separating, or increasing in size.

\begin{figure}
\centering
\includegraphics[width=0.5\textwidth]{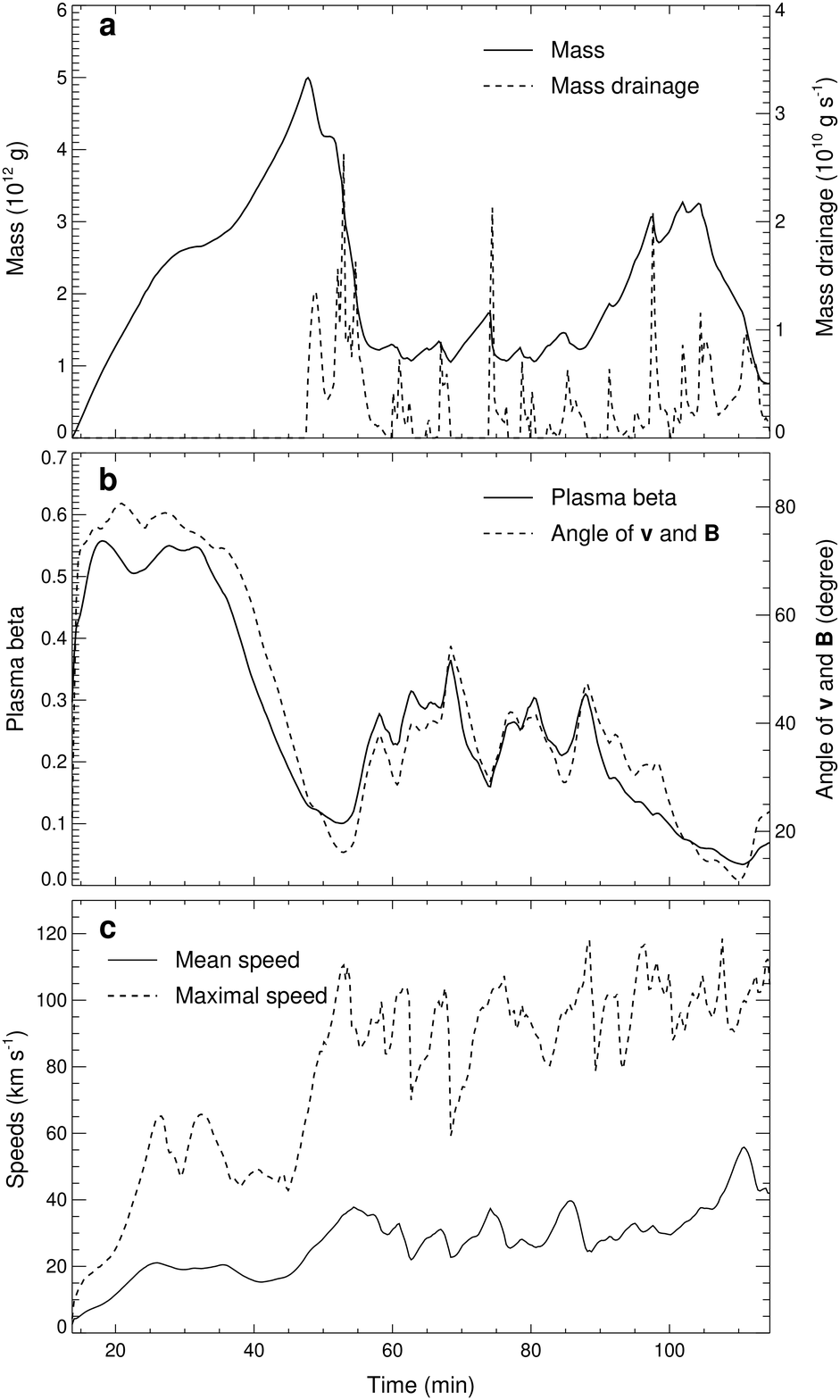}
\caption{Time evolution of coronal rain plasma properties shown as time curves of (a) total 
coronal rain mass (solid line) and mass drainage rate (dashed line); (b) mean plasma $\beta$
 (solid line) and the angle between local velocity and magnetic field vector (dashed line);  and
(c) instantaneous mean speed (solid line) and maximal speed (dashed line) of coronal rain blobs.}
\label{fig:tcurve}
\end{figure}

In order to understand the time evolution, we quantify several properties of the coronal rain 
blobs in terms of mass, mass drainage rate, mean plasma $\beta$, angle between local velocity and 
magnetic field vector, instantaneous mean speed and maximal speed, and plot the time evolution 
curves in Figure~\ref{fig:tcurve}. Coronal rain blobs are cool and dense plasma above transition 
region, so we practically define that they are composed of cells with density larger than 
$7\times10^{10}$ cm$^{-3}$, with temperature lower than 0.1 MK, and with height larger than 6 Mm. 
The mass drainage rate is the integrated coronal rain mass flux through a horizontal plane at 
6.5 Mm height. During the first condensation from
14 min to 30 min, the condensation has a roughly linear growth rate of $2.84\times10^{9}$ g s$^{-1}$,
which is much slower than the one in 3D prominence simulation ($1.14\times10^{10}$ g s$^{-1}$) 
\citep{Xia16} and the estimation from prominence observations ($1.2\times10^{10}$ g s$^{-1}$) 
\citep{Liu12}. Comparing with our previous 2D coronal rain simulations \citep{Fang15} (assuming 1000 km
integral depth in the third direction), the condensation rate in this 3D model is an order of 
magnitude faster. From 30 min to 38 min, the growth rate slows down because the fully
developed RTI breaks the initial large blobs into pieces in which tiny blobs and very thin
threads are heated to hotter and less dense material than coronal rain plasma mainly by thermal 
conduction. From 38 min until the first peak of the mass curve at 47.8 min, the growth rate is
about $3.25\times10^{9}$ g s$^{-1}$, slightly faster than in the initial phase. After the first peak,
the mass begins to drop rapidly when fast drainage due to falling of coronal rain starts. This 
fast drainage, which finishes at 60 min, is contributed by the large  falling clumps at the two flanks.
After that, smaller coronal rain blobs formed at the central loop parts stream down episodically. As 
continuous condensation increases the coronal rain mass, the episodic fallen blobs, which correspond
to the peaks in the mass drainage curve, decrease the mass. Therefore, the total mass remains at a 
steady level for a while from 55 min to 85 min. After that, the mass increases and decreases 
forming a large hill in the curve over a period of 25 min when a rain shower forms and falls. The plasma $\beta$ of the
condensation is relatively high (more than 0.5) at the beginning as the initial condensation 
forms in a weak magnetic field region at the loop top. With the solar gravity, RTI is 
inevitable under these circumstances and the nearly vertical falling of the condensation spikes
pushes the magnetic loops down, making the angle between velocity and magnetic field close to 80 
degrees. As the coronal rain blobs fall to lower regions with stronger magnetic field and smaller
plasma $\beta$, they are ultimately forced to follow the guide of magnetic field lines with the angle
decreasing. Therefore, the curve of plasma $\beta$ and the angle match each other as shown in 
Figure~\ref{fig:tcurve} (b). The mean speed of coronal rain plasma at each instant shifts from 
about 20 km s$^{-1}$ to about 30 km s$^{-1}$ after the fall of the large flank blobs, while the
instantaneous maximal speed, which is about two  to three times larger than the instantaneous mean 
speed, also increases to about 100 km s$^{-1}$. 

\begin{figure}
\centering
\includegraphics[width=0.5\textwidth]{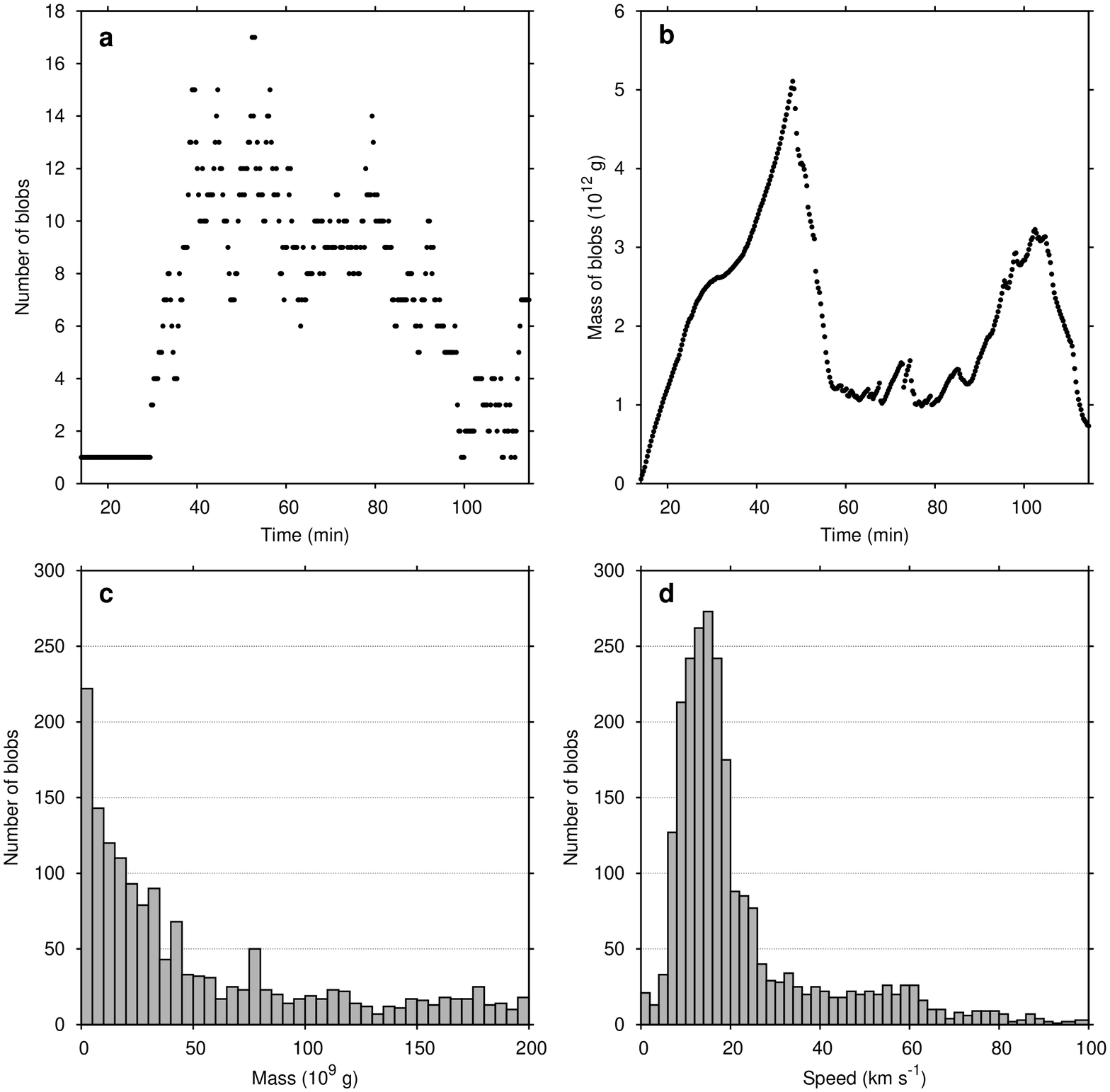}
\caption{Statistic analysis of coronal rain blobs. (a) the number of coronal rain blobs at each 
instant; (b) total mass of coronal rain blobs; (c) distribution function of mass of blobs in all
snapshots; and (d) distribution function of speed of blobs in all snapshots.}
\label{fig:statistic}
\end{figure}

We wrote a program to automatically count the coronal rain blobs in each snapshot and quantify 
their properties, such as number, mass, volume, density, centroid location, and centroid 
speed \citep[see][for details]{Moschou15}. The key algorithm is to group adjacent cells that contain
coronal rain plasma into one blob, above TR, which is detected based on 
the vertical gradient of local density and temperature. To consider fully resolved blobs, we neglect tiny 
blobs containing less than 64 cells at the highest refinement level. Using this program, we performed statistical analysis
of coronal rain blobs as shown in Figure~\ref{fig:statistic}. The total number of blobs in all 352 
snapshots is 2406. The mean number of blobs per snapshot is 6.8. The maximal number of blobs in one snapshot
is 17. In the curve of blob number, we can see that the initial big blob breaks into three pieces at 30 min. 
Then the number increases up to 17 at 55 min and decreases in the later phase. The total mass of coronal 
rain blobs matches well with the mass curve in Figure~\ref{fig:tcurve} (a). Blobs are dominated by small ones
with mass less than $5\times10^{10}$ g with the heaviest blob of $3.2\times10^{12}$ g. The distribution of 
speed has a peak around 15 km s$^{-1}$ and a long tail over 100 km s$^{-1}$. The mean speed of all blobs is
22.4 km s$^{-1}$. The blob with the maximal speed of 104.5 km s$^{-1}$ is very small with a mass of 
$2.4\times10^{9}$ g.

\begin{figure}
\centering
\includegraphics[width=0.5\textwidth]{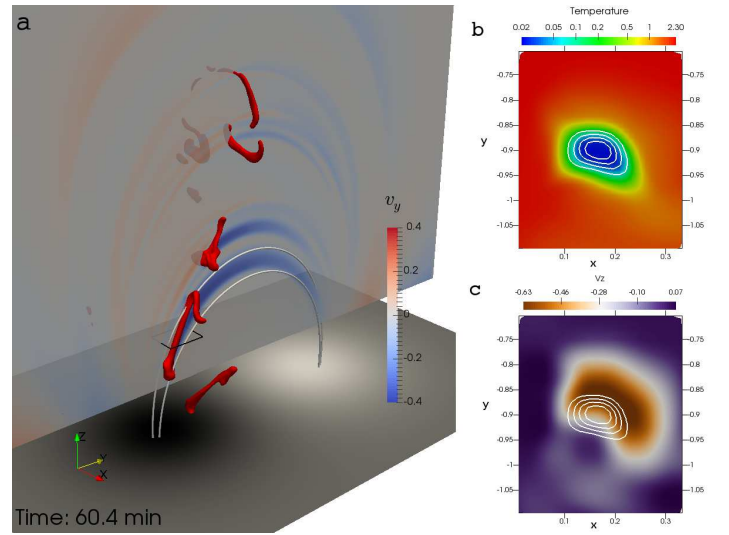}
\caption{Case study of a stretched coronal rain blob at 60.4 min. (a) shows a 3D view of blobs in red 
isosurface of density, two magnetic field lines going through the head and tail of a blob, 
a vertical plane cutting through the blobs with translucent color indicating $v_y$; (b)  shows the 
horizontal slice through the blob bounded by the black frame in (a) colored by temperature with
number density contours in values of 7, 11, 15, 19, 23 $n_0$; and (c) shows the slice colored by $v_z$ 
with the same density contours as (b).}
\label{fig:fallblob}
\end{figure}

After analyzing the overall properties of the modeled coronal rain event, we go into 
more detail with case studies of typical coronal rain blobs. In Figure~\ref{fig:fallblob}, we
find several coronal rain blobs, presented by red density isosurfaces, at different heights in 
different time stages of their life. We focus on a stretched coronal rain blob in the lower left part
of panel (a). It is falling roughly in the central plane with $x$ close to 0. It has a 
swelled head with relatively high speed (50 km s$^{-1}$) before hitting the TR, a
straight elongated body, and a hook-like tail. The blob is falling along a bundle of magnetic
loops with its lower parts residing on the shorter loops, as shown by two field lines
going through the head and the tail part. 
A vertical slice cutting through the blobs shows the $v_y$ distribution. There are 
fast following flows in the wake of the falling blob. The elongating blob has low gas pressure in
its body inducing siphon flows that enhance the wake flows (see the blue regions). The siphon flows  upstream of the falling blob become neutralized by the pushing motion of the blob. Blobs near the
loop top in higher regions have little motion along the loops, so the siphon flows from both sides
towards the blobs are obvious and bring material to feed the growth of the blobs. Once the blobs 
reach the low-$\beta$ region, the lower part at the lower loops feels stronger projected gravity due to
the shape of the loops. The lower part falls faster than the upper part along the loops, so an 
initially compact blob gets elongated and this tearing also decreases the gas pressure in the 
straight body of an elongated blob. Besides, when a plasma blob streams down in a flux tube with
narrower cross-section closer to the footpoints, it will be squeezed and elongated.
We made a small horizontal slice cutting across the selected blob, 
and plot the temperature map (panel (b)) and $v_z$ map (panel (c)) with number density contours on 
them. The blob is wrapped by a TR which connects the 0.04 MK coronal rain plasma
to 1 MK coronal plasma in a range of about 500 km. The thickness of the blob in this cross-section is 
about 700 km if we count the 7 $n_0$ density contour and 0.1 MK temperature contour as its border. The
density contours and the temperature map show increasing density and decreasing temperature when
getting closer to the center of the blob, which reflects the multi-density and multi-thermal 
nature of the coronal rain blobs \citep{Antolin15}. In the vertical velocity map, the downflow 
region occupies not only the coronal periphery but also half the blob itself. These anti-parallel 
shearing flows are divided over the core of the blob, which leads to elongation of the blob during 
its fall. The large velocity difference between different parts of individual blobs is also 
reflected in the significant difference between the maximal speed and the mean speed at each 
snapshot in Figure~\ref{fig:tcurve} (c).

\begin{figure}
\centering
\includegraphics[width=0.5\textwidth]{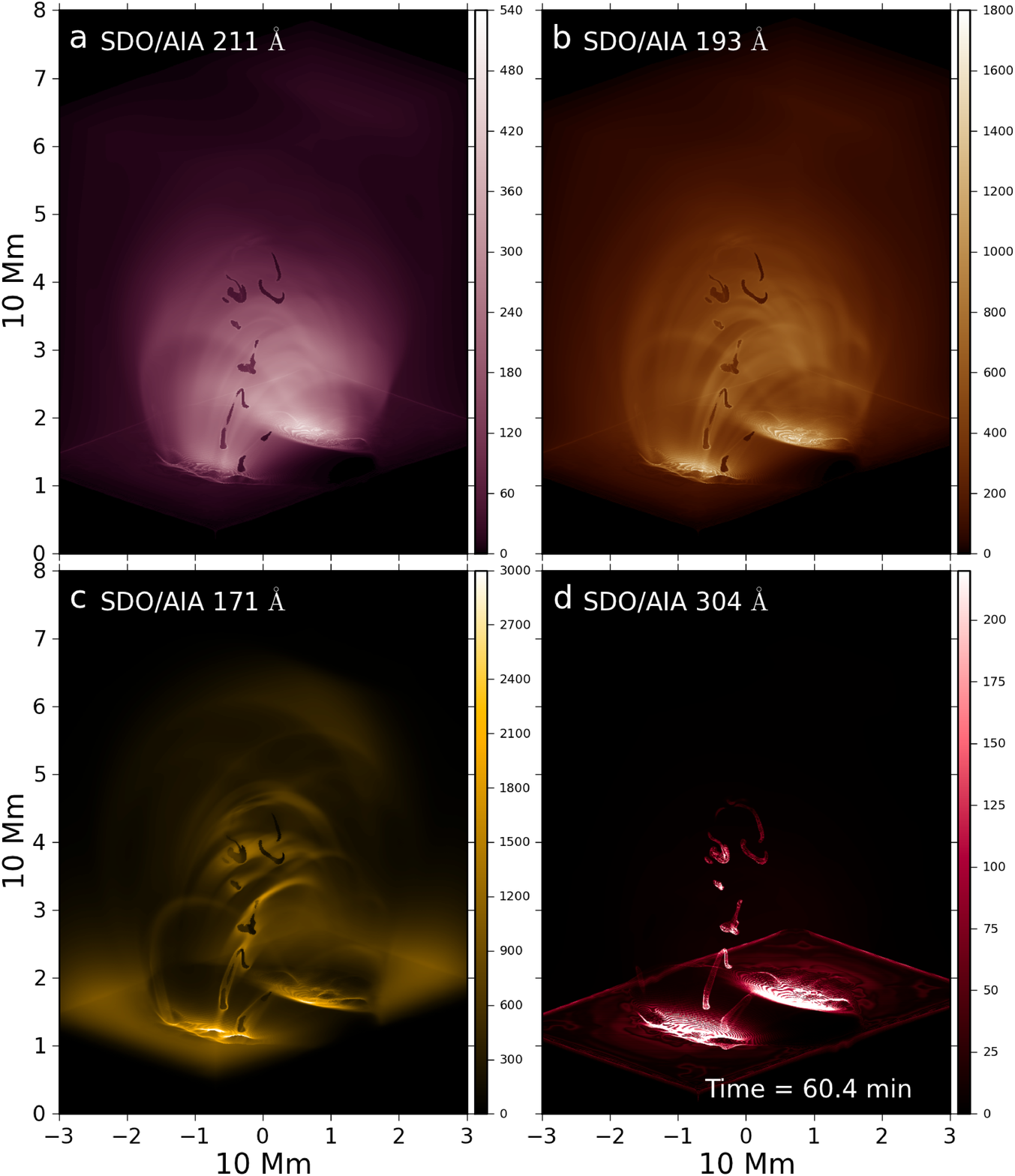}
\caption{SDO/AIA synthetic views on coronal rain blobs at 60.4 min in (a) 211~\AA~; (b)
193~\AA~; (c) 171~\AA~; and (d) 304~\AA~ wave channels. Numbers in the color bars have a unit
of DN s$^{-1}$. Supplementary Movie 2 showing this figure is available online.}
\label{fig:fsyn}
\end{figure}

To directly compare with observed images from the Solar Dynamics Observatory Atmospheric 
Imaging Assembly (AIA) instrument, we made synthetic observations on our simulations in 
selected viewing angle and retrieved synthetic images with a technique described in detail in  \citet{Xia14}.
To mimic absorption of EUV emission by the core plasma of condensations, we exclude emission 
coming from behind dense plasma with a number density larger than $2\times10^{10}$ cm$^{-3}$,
except for the 304~\AA~ wave channel.
For example, Figure~\ref{fig:fsyn} shows synthetic views on the modeled coronal rain and 
coronal loops at 60.4 min in 211~\AA, 193~\AA, 171~\AA, and 304~\AA~ wave channels, which have
main contribution temperatures of about 1.8, 1.5, 0.8, and 0.08 MK, respectively. The line
 of sight of Figure~\ref{fig:fsyn} has an azimuthal angle of 45$^{\circ}$ away from the 
PIL and an elevation angle of 20$^{\circ}$. In hot channels panel (a), (b), and (c), 
the bright coronal loops and their footpoint regions appear more dispersed in higher-temperature channels. Dark clumps of coronal rain reside in dim loops in 211~\AA, while 
adhering to bright halo segments of 171~\AA~ loops. Bright clumps of coronal rain are found
in the cooler 304~\AA~ channel sampling TR plasma. The impact site of coronal rain at the
chromosphere is at this moment a fairly compact region of negative polarity with a diameter 
less than 10 Mm. In the movie of Figure~\ref{fig:fsyn}, the inital condensation is 
barely seen in 304~\AA~, because the density of the emission-contributing plasma 
(around 0.08 MK) is relatively low (about $2\times10^9$ cm$^{-3}$) at the transition layer 
between the condensation and corona. 

\begin{figure*}
\centering
\includegraphics[width=0.8\textwidth]{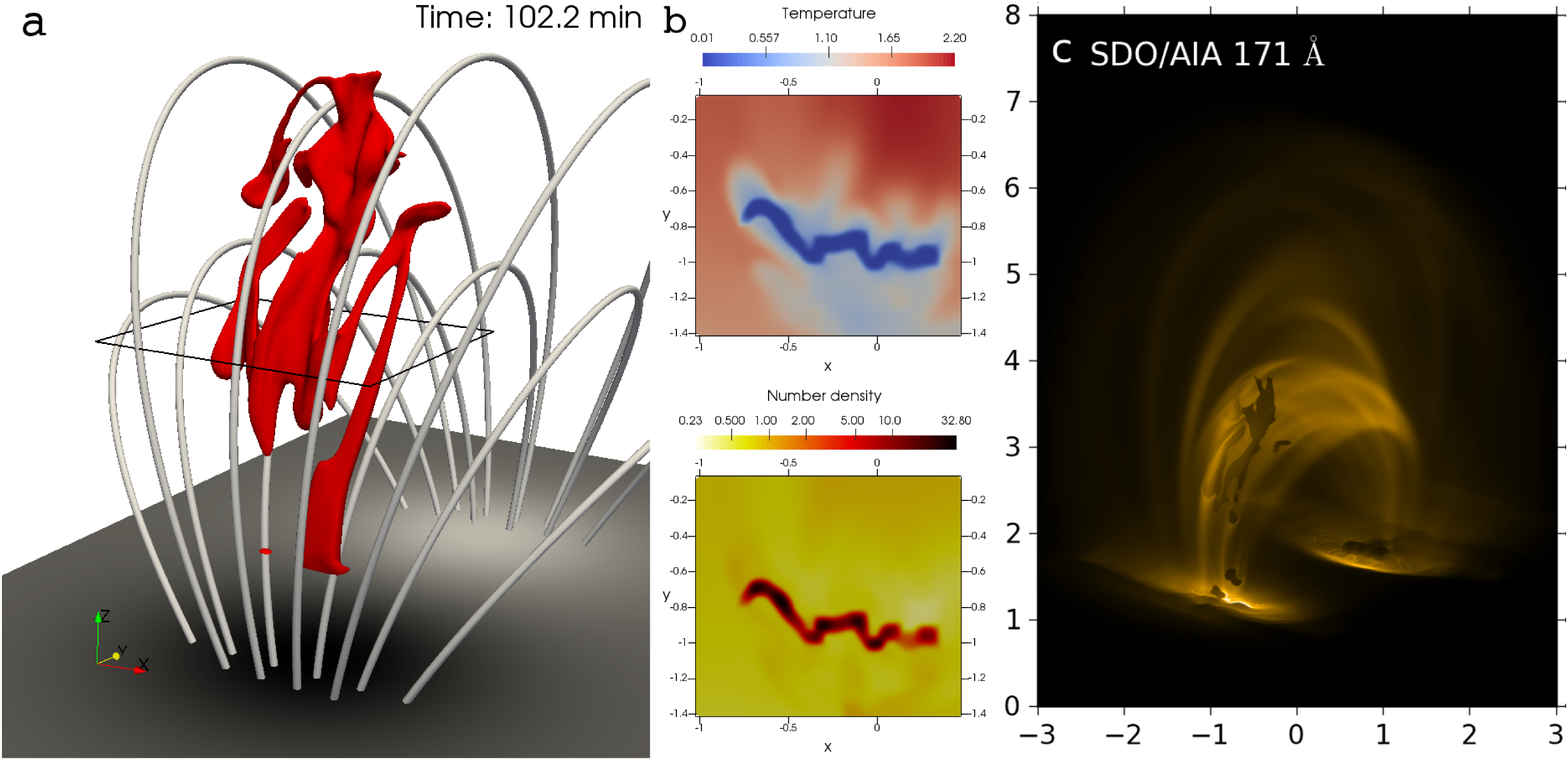}
\caption{Coronal rain ``shower'' at 102.2 min. (a) shows a 3D view with red density isosurfaces and selected
magnetic field lines; (b) shows a plane slice across the rain blobs, indicated by the black frame in (a),
showing temperature map and number density map; and (c) shows a synthetic view in AIA 171~\AA~ wave channel.}
\label{fig:fshower}
\end{figure*}

In the last phase of the simulation from 90 min to 113 min, a bursty condensation makes a 
coronal rain ``shower'' in a large bundle of coronal loops. The total mass of the condensation 
increases and peaks at 105 min. The total number of blobs first increases and soon 
decreases because neighboring small condensations grow fast enough to touch one another and 
to be counted as one blob. We plot a representative snapshot of this phase in Figure~\ref{fig:fshower}.
As shown by density isosurfaces in panel (a), long streams of condensed plasma are connected 
to one another. The horizontal extension of the streams along the $x$-direction, covering loops 
with similar length, is about 10 Mm, while the thickness of the streams along $y$-direction 
is around 1 Mm. In panel (b), a horizontal cross-section of the streams finds the density 
inhomogeneities within the coronal rain clumps, which is similar to the results found in 
observations \citep{Antolin15}, although the spatial sizes are different. In our model,
a coronal rain clump can have density inhomogeneity because it is composed of several seed blobs 
which grow and move along the magnetic loops closely in space and time. In panel (c), a synthetic view in AIA 
171~\AA~ channel shows dark clumps, which are the high-density ($2\times10^{10}$ cm$^{-3}$) 
cores of coronal rain streams in panel (a), in bright coronal loops.

\section{Conclusions and discussion}\label{sec:conc}

We realized a 3D coronal rain simulation with unprecedented high resolution and a more 
realistic magnetic environment than previous models. The formation process of coronal rain, 
characterized by progressive cooling and condensation in situ, is reproduced in our model.
The physical reason behind it is the radiation-dominated thermal instability. The formation
of the first condensation is presented in detail with its 3D shape extending along the PIL and with
the shape of rebounding shock fronts indicating the time sequence of the condensation progress. The
first long condensation on the loop top region develops RTI interchange deformation and becomes 
fragmented into smaller blobs. The blobs fall slowly and vertically against magnetic field loops
until reaching low-$\beta$ regions and start to slip and fall along the field loops from loop top to
loop footpoints. A similar phenomenon was presented by \citet{Petralia16} in a 3D MHD simulation of
the falling of dense fragments along magnetic 
flux tubes of an erupted solar filament, in which the initial falling speed 
of the dense blobs is about 300 km s$^{-1}$, much higher than our case, and fast enough to 
generate slow shocks, which are not found ahead of the falling blobs in our study. They also 
found that these fast-moving blobs with imposed initial motions misaligned with magnetic 
field soon become deformed and mixed due to the feedback from dragged magnetic field lines 
\citep{Petralia17}. Similarly, the misaligned motion of our falling blobs near the loop tops
 also leads to reshaping and fragmentation of the blobs.

Statistical study of the coronal rain blobs found in all snapshots in our simulation shows that
small blobs with mass less than $10^{10}$ g dominate the population despite the fact  that unusually large
 blobs show up in the initial and last phases. Large blobs have statistically slower centroid 
speed than small ones. When the blobs stream along the magnetic loops, they are stretched and 
develop a non-uniform velocity pattern. The elongated blobs have faster (siphon) flow on those 
parts closer to the center of curvature of the hosting coronal loops. AIA synthetic observations of our 
simulated coronal rain well resemble real observations in several aspects, like dark falling clumps 
in hot EUV channels and bright rain blobs in cool 304~\AA~ channel. We also find density
inhomogeneities within the coronal rain clumps in a rain ``shower'', which may explain the 
observed multi-stranded nature of coronal rain \citep{Antolin15}.

The magnetic field strength in this model is as yet relatively low compared to typical active region
coronal loops at the same height. It is computationally challenging to simulate plasma 
with extremely low $\beta$ and high Alfv\'en speed when solving conservative energy equations and 
using explicit schemes. At the cost of much more computational resources, we can get
closer to the realistic magnetic field strength in the work in progress. The RTI interchange mode of
loop-top condensations may then be suppressed by the rigid line-tied coronal loops in very 
low-$\beta$ regimes. The physics of rain-blob formation and the dynamics of coronal rain
 blobs in the lower regions should remain the same. Since the RTI interchange mode of prominence plasma
has been observed \citep{Berger08,Berger10} and modeled \citep{Hillier11,Keppens15,Xia16b} in the 
quiescent region, it may be found on high-$\beta$ loop-top by future observations. Previous 
simulations of RTI in prominences start with existing static prominence material excluding the dynamic 
formation process; the model here is the first one covering both formation and subsequent RTI of 
prominence-like plasma in similar local magnetic environment, for example, weak and locally horizontal magnetic 
field. However, the vertical thickness of the initial condensation is only about 2 Mm, much smaller
than a typical prominence and the bipolar magnetic structure is quite different from prominence-hosting
magnetic structures with magnetic dips, thus the details of RTI here are different from those in 
modeled prominences and observed prominences. We further plan to use 
realistic magnetic field extrapolated from actual observed magnetograms and study the dynamics of coronal
rain blobs. Flare-driven coronal rain has not yet been simulated and needs simulations for further 
understanding of its fine-scale strands and the cooling of post-flare loops.

\begin{acknowledgements}
C.X. wishes to thank FWO (Research Foundation Flanders) for the award of 
postdoctoral fellowship. C.X. appreciates the suggestive discussion with Dr. Patrick Antolin.
 This research was supported by FWO and by KU Leuven Project No. GOA/2015-014 and by the 
Interuniversity Attraction Poles Programme by the Belgian Science Policy Office (IAP P7/08 
CHARM). The simulations were conducted on the VSC (Flemish Supercomputer 
Center funded by Hercules foundation and Flemish government). 
\end{acknowledgements}

\bibliographystyle{aa}
\bibliography{refads}

\end{document}